\documentclass[12pt]{article}
\usepackage{graphicx}
\usepackage{color}

\def\hybrid{\topmargin 0pt      \oddsidemargin 0pt
        \headheight 0pt \headsep 0pt
       \voffset-1cm
        \textwidth 6.25in       
       \textheight 9.5in       
        \marginparwidth 0.0in
        \parskip 5pt plus 1pt   \jot = 1.5ex}
\catcode`\@=11
\def\marginnote#1{}

\newcount\hour
\newcount\minute
\newtoks\amorpm
\hour=\time\divide\hour by60
\minute=\time{\multiply\hour by60 \global\advance\minute by-\hour}
\edef\standardtime{{\ifnum\hour<12 \global\amorpm={am}%
        \else\global\amorpm={pm}\advance\hour by-12 \fi
        \ifnum\hour=0 \hour=12 \fi
        \number\hour:\ifnum\minute<10 0\fi\number\minute\the\amorpm}}
\edef\militarytime{\number\hour:\ifnum\minute<10 0\fi\number\minute}

\def\draftlabel#1{{\@bsphack\if@filesw {\let\thepage\relax
   \xdef\@gtempa{\write\@auxout{\string
      \newlabel{#1}{{\@currentlabel}{\thepage}}}}}\@gtempa
   \if@nobreak \ifvmode\nobreak\fi\fi\fi\@esphack}
        \gdef\@eqnlabel{#1}}
\def\@eqnlabel{}
\def\@vacuum{}
\def\draftmarginnote#1{\marginpar{\raggedright\scriptsize\tt#1}}

\def\draftlabel#1{{\@bsphack\if@filesw {\let\thepage\relax
   \xdef\@gtempa{\write\@auxout{\string
      \newlabel{#1}{{\@currentlabel}{\thepage}}}}}\@gtempa
   \if@nobreak \ifvmode\nobreak\fi\fi\fi\@esphack}
        \gdef\@eqnlabel{#1}}
\def\@eqnlabel{}
\def\@vacuum{}
\def\draftmarginnote#1{\marginpar{\raggedright\scriptsize\tt#1}}

\def\draft{\oddsidemargin -.5truein
        \def\@oddfoot{\sl preliminary draft \hfil
        \rm\thepage\hfil\sl\today\quad\militarytime}
        \let\@evenfoot\@oddfoot \overfullrule 3pt
        \let\label=\draftlabel
        \let\marginnote=\draftmarginnote
   \def\@eqnnum{(\theequation)\rlap{\kern\marginparsep\tt\@eqnlabel}%
\global\let\@eqnlabel\@vacuum}  }

\def\numberbysection{\@addtoreset{equation}{section}
        \def\theequation{\thesection.\arabic{equation}}}

\def\underline#1{\relax\ifmmode\@@underline#1\else
        $\@@underline{\hbox{#1}}$\relax\fi}

\def\titlepage{\@restonecolfalse\if@twocolumn\@restonecoltrue\onecolumn
     \else \newpage \fi \thispagestyle{empty}\c@page\z@
        \def\thefootnote{\fnsymbol{footnote}} }

\def\endtitlepage{\if@restonecol\twocolumn \else  \fi
        \def\thefootnote{\arabic{footnote}}
        \setcounter{footnote}{0}}  
\relax

\hybrid

\newfont{\Bbb}{msbm10 scaled 1\@ptsize00}
\newfont{\Bbbb}{msbm7 scaled 1\@ptsize00}

\newcommand{\DDD}{\raise-1pt\hbox{$\mbox{\Bbbb D}$}}

\newcommand{\UUU}{\raise-1pt\hbox{$\mbox{\Bbbb U}$}}

\newcommand{\z}{\raise-1pt\hbox{$\mbox{\Bbbb Z}$}}

\def\beq{\begin{equation}}
\def\eeq{\end{equation}}
\def\p{\partial}

\begin{document}

\begin{titlepage}

\title{Dynamics of poles of elliptic solutions to the BKP equation}

\author{D.~Rudneva\thanks{National Research University Higher School of Economics,
20 Myasnitskaya Ulitsa, Moscow 101000, Russian Federation;
Skolkovo Institute of Science and Technology, 143026 Moscow, Russian Federation; \newline
e-mail: missdaryarudneva@gmail.com
}
\and
 A.~Zabrodin\thanks{National Research University Higher School of Economics,
20 Myasnitskaya Ulitsa, Moscow 101000, Russian Federation;
ITEP, 25 B.Cheremushkinskaya, Moscow 117218, Russian Federation;
Skolkovo Institute of Science and Technology, 143026 Moscow, Russian Federation;
e-mail: zabrodin@itep.ru}
}

\maketitle

\vspace{-7cm} \centerline{ \hfill ITEP-TH-26/18}\vspace{7cm}

\begin{abstract}

We derive equations of motion for poles of elliptic solutions to the 
B-version of the Kadomtsev-Petviashvili equation (BKP).
The basic tool is the auxiliary linear problem for the Baker-Akhiezer function.
We also discuss integrals of motion for the pole dynamics which follow from the equation
of the spectral curve.

\end{abstract}

\end{titlepage}

\vspace{5mm}

\section{Introduction}

In the seminal 
paper \cite{AMM77} the motion of poles of singular solutions to the
Korteweg-de Vries and Boussinesq equations was investigated. 
It was discovered that the poles move as particles of the 
many-body Calogero-Moser system \cite{Calogero71,Calogero75,Moser75} 
with some additional restrictions in the phase space.
In \cite{Krichever78,CC77} it was shown that in the case of the Kadomtsev-Petviashvili (KP)
equation this correspondence becomes an isomorphism: the dynamics of 
poles of rational solutions to the
KP equation is given by equations of motion for the Calogero-Moser system 
with pairwise interaction potential
$1/(x_i-x_j)^2$. This remarkable connection
was further generalized to elliptic (double periodic) 
solutions by Krichever in \cite{Krichever80}:
poles $x_i$ of the elliptic solutions move according to the equations of motion
\beq\label{int1}
\ddot x_i=4\sum_{k\neq i} \wp ' (x_i-x_k)
\eeq
of Calogero-Moser particles with the elliptic
interaction potential $\wp (x_i-x_j)$ ($\wp$ is the Weierstrass $\wp$-function).
This many-body system of classical mechanics is known to be integrable. 
For a review of the models of the Calogero-Moser type (including the models associated 
with the classical root systems) see \cite{OP81}. (For further progress, generalizations
and related models see, e.g., \cite{GH84}--\cite{Ch19}.) 

This result allows for generalizations in various directions. 
The extension to the matrix KP equation was discussed in \cite{KBBT95}; in this case 
the poles and matrix residues at the poles move as particles of the spin generalization
of the Calogero-Moser model known also as the Gibbons-Hermsen model \cite{GH84}.
Another generalization of the 
Calogero-Moser many-body systems with elliptic interaction is their relativistic 
extension known also as the Ruijsenaars-Schneider systems \cite{RS86,Ruij87}
and their versions with spin degrees of freedom \cite{KZ95}. These relativistic
systems emerge as dynamics of poles of elliptic solutions to the two-dimensional
Toda lattice (see \cite{KZ95}). 

We are going to suggest a generalization 
in yet another direction: our goal is to find out what dynamical system governs
the dynamics of poles of elliptic solutions to the B-version of the KP equation. 
We will see that the result is a new, previously unknown dynamical system 
with elliptic interaction which does
not look like any kind of the Calogero-Moser system.

The method of derivation of 
equations of motion for the poles of singular solutions to integrable non-linear
equations
suggested by Krichever consists in substituting the pole ansatz not in the 
non-linear equation itself
but in the auxiliary linear problems for it. We apply this method
in this paper. 

In Section 2 
we derive equations of motion for poles of elliptic solutions to the 
$B$-version of the KP equation (BKP). The BKP equation is the first member of an 
infinite BKP hierarchy with independent variables (``times'') 
$t_1, t_3, t_5, t_7, \ldots$ \cite{DJKM83,DJKM82}, see also \cite{DJKM82a,LW99,Tu07}. 
We set $t_1=x$.
The BKP equation has the form of a system of two partial differential
equations for two dependent variables $u$, $w$:
\beq\label{int2}
\left \{
\begin{array}{l}
3w' =10 u_{t_3}+20 u^{'''} +120 uu'
\\ \\
w_{t_3}-6u_{t_5}=w^{'''}-6u^{'''''}-60 uu^{'''}-60 u'u'' +6uw'-6wu',
\end{array}
\right.
\eeq
where prime means differentiation w.r.t. $x$. In fact the variable $w$ can be excluded 
and the equation can be written in terms a single dependent variable $U=\int^x udx$. 
Equations (\ref{int2}) are equivalent to the Zakharov-Shabat (``zero curvature'') equation 
$\p_{t_5}B_3-\p_{t_3}B_5+[B_3, B_5]=0$ for the differential operators
\beq\label{int3}
B_3=\p_x^3+6u\p_x, \qquad B_5=\p_x^5+10u\p_x^3+10u'\p_x^2+w\p_x.
\eeq
In its turn, the Zakharov-Shabat equation is the compatibility condition for 
the auxiliary linear problems
$$
\p_{t_3}\psi =B_3\psi , \qquad \p_{t_5}\psi =B_5\psi
$$
for the Baker-Akhiezer function $\psi$ which depends on a spectral parameter $z$.

The change of dependent variables from $u,w$ to the tau-function 
$\tau = \tau (x, t_3, t_5, \ldots )$
\beq\label{int4}
u=\p_x^2\log \tau , \qquad w=\frac{10}{3}\, \p_{t_3}\p_x \log \tau +
\frac{20}{3}\, \p_x^4 \log \tau +20 (\p_x^2 \log \tau )^2
\eeq
makes the first of the equations (\ref{int2}) trivial 
and the other one turns into the bilinear form \cite{DJKM82}
\beq\label{int5}
\Bigl (D_1^6 -5D_1^3D_3-5D_3^2+9D_1D_5 \Bigr ) \tau \cdot \tau =0,
\eeq
where $D_i$ are the Hirota operators. Their action is defined by the rule
$$
P(D_1, D_3, D_5, \ldots )\tau \cdot \tau =
P(\p_{y_1}, \p_{y_3}, \p_{y_5}, \ldots )
\tau (x+y_1, t_3+y_3, \ldots )\tau (x-y_1, t_3-y_3, \ldots )\Bigr |_{y_i=0}
$$
for any polynomial $P(D_1, D_3, D_5, \ldots )$. The Baker-Akhiezer function
is known to be expressed through the tau-function according to the formula
\beq\label{psi}
\psi = A(z)\exp \, \Bigl (\, \sum_{k\geq 1, \, k \,\, {\rm odd}}t_k z^k \Bigr )\,
\frac{\tau \Bigr (t_1 -2z^{-1}, t_3-\frac{2}{3}\, z^{-3}, t_5-\frac{2}{5}\, z^{-5},
\ldots \Bigr )}{\tau (t_1, t_3, t_5, \ldots )}.
\eeq
Here $A(z)$ is the normalization factor.

Our aim is to study double-periodic (elliptic) in the variable $t_1=x$ 
solutions of the BKP equation. For such solutions the tau-function is an 
``elliptic polynomial'' in the variable $x$:
\beq\label{int6}
\tau = A e^{cx^2/2}\prod_{i=1}^{N}\sigma (x-x_i)
\eeq
with some constants $A, c$, where 
$$
\sigma (x)=\sigma (x |\, \omega , \omega ')=
x\prod_{s\neq 0}\Bigl (1-\frac{x}{s}\Bigr )\, e^{\frac{x}{s}+\frac{x^2}{2s^2}},
\quad s=2\omega m+2\omega ' m' \quad \mbox{with integer $m, m'$},
$$ 
is the Weierstrass 
$\sigma$-function with quasi-periods $2\omega$, $2\omega '$ such that 
${\rm Im} (\omega '/ \omega )>0$. It is connected with the Weierstrass 
$\zeta$- and $\wp$-functions by the formulas $\zeta (x)=\sigma '(x)/\sigma (x)$,
$\wp (x)=-\zeta '(x)=-\p_x^2\log \sigma (x)$.
The roots $x_i$ are assumed to be 
all distinct. Correspondingly, 
the function $u=\p_x^2\log \tau$ is an 
elliptic function with double poles at the points $x_i$:
\beq\label{int7}
u=c-\sum_{i=1}^{N}\wp (x-x_i).
\eeq
The poles depend on the times $t_3$, $t_5$. We will show that the dependence 
on the time $t_3=t$ is described by the equations of motion
\beq\label{int8}
\ddot x_i +6\sum_{j\neq i}(\dot x_i +\dot x_j)\wp '(x_i-x_j)-72\!\!
\sum_{j\neq k \neq i}\wp (x_i-x_j)\wp '(x_i-x_k)=0.
\eeq
This is the main result of the paper. A characteristic feature of the system 
(\ref{int8}) is the presence of both two-body and three-body interaction and 
dependence on the first time derivatives. Note that the latter is also the case for the 
relativistic Calogero-Moser systems \cite{RS86,Ruij87} while the former seems to 
be a novel phenomenon for classical integrable systems.

In Section 3 we discuss integrals of motion for the dynamical system 
(\ref{int8}). It is shown that there is a large set of integrals of motion.
In Section 4 properties of the spectral curve are studied. Section 5 is devoted
to analytic properties of the $\psi$-function on the spectral curve.

\section{Elliptic solutions to the BKP equation and dynamics of poles}

According to Krichever's method \cite{Krichever80}, the basic tool for 
studying $t$-dynamics of poles is the auxiliary linear problem 
$\p_{t}\psi =B_3\psi$ for the 
function $\psi$, i.e.,
\beq\label{ba0}
\p_t \psi =\p_x^3\psi +6u \p_x \psi .
\eeq
Since the coefficient function $u$ is double-periodic, 
one can find double-Bloch solutions $\psi (x)$, i.e., solutions such that 
$\psi (x+2\omega )=b \psi (x)$, $\psi (x+2\omega ' )=b' \psi (x)$
with some Bloch multipliers $b, b'$. Equations (\ref{psi}), (\ref{int6}) tell us that the
Baker-Akhiezer function has simple poles at the points $x_i$.
The pole ansatz for the $\psi$-function is
\beq\label{ba1}
\psi = e^{xz+tz^3}\sum_{i=1}^N c_i \Phi (x-x_i, \lambda ),
\eeq
where the coefficients $c_i$ do not depend on $x$.
Here we use the function
$$
\Phi (x, \lambda )=\frac{\sigma (x+\lambda )}{\sigma (\lambda )\sigma (x)}\,
e^{-\zeta (\lambda )x}
$$
which has a simple pole
at $x=0$ ($\zeta$ is the Weierstrass $\zeta$-function). 
The expansion of $\Phi$ as $x\to 0$ is
$$
\Phi (x, \lambda )=\frac{1}{x}+\alpha_1 x +\alpha_2 x^2 +\ldots , \qquad 
x\to 0,
$$
where $\alpha_1=-\frac{1}{2}\, \wp (\lambda )$, $\alpha_2=-\frac{1}{6}\, \wp '(\lambda )$. 
The parameters $z$ and $\lambda$ 
are spectral parameters, they are going to be connected by equation of the spectral curve.
Using the quasiperiodicity properties of the function $\Phi$,
$$
\Phi (x+2\omega , \lambda )=e^{2(\zeta (\omega )\lambda - \zeta (\lambda )\omega )}
\Phi (x, \lambda ),
$$
$$
\Phi (x+2\omega ' , \lambda )=e^{2(\zeta (\omega ' )\lambda - \zeta (\lambda )\omega ' )}
\Phi (x, \lambda ),
$$
one can see that $\psi$ given by (\ref{ba1}) 
is indeed a double-Bloch function with Bloch multipliers
$$b=e^{2(\omega z + \zeta (\omega )\lambda - \zeta (\lambda )\omega )}, \qquad
b '=e^{2(\omega ' z + \zeta (\omega ' )\lambda - \zeta (\lambda )\omega ' )}.$$
We will often suppress the second argument of $\Phi$ writing simply 
$\Phi (x)=\Phi (x, \lambda )$. 
We will also need the $x$-derivatives  
$\Phi '(x, \lambda )=\p_x \Phi (x, \lambda )$, 
$\Phi ''(x, \lambda )=\p^2_x \Phi (x, \lambda )$,
etc. 

It is evident from (\ref{int7}) and (\ref{ba0}) that the constant $c$ 
in the pole expansion for the function $u$ can be eliminated
by the simple 
transformation $x\to x-6ct$, $t\to t$ (or $\p_x \to \p_x$, $\p_t \to \p_t +6c\p_x$
for the vector fields). Because of this we will put $c=0$ from now on for simplicity.

Substituting (\ref{ba1}) into (\ref{ba0}) with  
$\displaystyle{u=-\sum_{i}\wp (x-x_i)}$, 
we get:
$$
\sum_i \dot c_i\Phi (x-x_i)-\sum_i c_i \dot x_i \Phi '(x-x_i)=3z^2
\sum_i c_i \Phi '(x-x_i)+3z\sum_i c_i \Phi ''(x-x_i)+\sum_i c_i \Phi '''(x-x_i)
$$
$$
-6z\Bigl (\sum_k \wp (x-x_k)\Bigr ) \Bigl (\sum_i c_i \Phi (x-x_i)\Bigr )
-6\Bigl (\sum_k \wp (x-x_k)\Bigr ) \Bigl (\sum_i c_i \Phi ' (x-x_i)\Bigr ).
$$
It is enough to cancel all poles which are at the points $x_i$ (up to fourth order).
It is easy to see that poles of the fourth and third order cancel identically. 
A direct calculation shows that the conditions of cancellation of second and first 
order poles have the form
\beq\label{ba2}
c_i\dot x_i=-(3z^2+6\alpha_1)c_i -6z \sum_{k\neq i}c_k \Phi (x_i-x_k)-6
\sum_{k\neq i}c_k \Phi '(x_i-x_k)+6c_i \sum_{k\neq i}\wp (x_i-x_k),
\eeq
\beq\label{ba3}
\begin{array}{lll}
\dot c_i &=&\displaystyle{
-6z\alpha_1 c_i -12\alpha_2c_i-6z\sum_{k\neq i}c_k \Phi '(x_i-x_k)
-6zc_i\sum_{k\neq i}\wp (x_i-x_k)}
\\ &&\\
&&\displaystyle{-\, 6 \sum_{k\neq i}c_k \Phi ''(x_i-x_k)
+6c_i \sum_{k\neq i}\wp '(x_i-x_k)}
\end{array}
\eeq
which have to be valid for all $i=1, \ldots , N$.
These conditions can be rewritten in the matrix form as
linear problems for a vector ${\bf c} =(c_1, \ldots , c_N)^T$:
\beq\label{a1}
\left \{ \begin{array}{l}
L{\bf c} = (3z^2 +6\alpha_1){\bf c}
\\ \\
\dot {\bf c} =M{\bf c},
\end{array}
\right.
\eeq
where
\beq\label{a1a}
L=-\dot X -6zA -6B +6D,
\eeq
\beq\label{a1b}
M= -(6z\alpha_1 +12\alpha_2)I -6zB -6zD -6C +6D'
\eeq
and the matrices $X$, $A$, $B$, $C$, $D$, $D'$, $I$ are given by
$X_{ik}=\delta_{ik}x_i$, $I_{ik}=\delta_{ik}$, 
$$
A_{ik}=(1-\delta_{ik})\Phi (x_i-x_k),
$$
$$
B_{ik}=(1-\delta_{ik})\Phi ' (x_i-x_k),
$$
$$
C_{ik}=(1-\delta_{ik})\Phi '' (x_i-x_k),
$$
$$
D_{ik}=\delta_{ik}\sum_{j\neq i}\wp (x_i-x_j),
$$
$$
D'_{ik}=\delta_{ik}\sum_{j\neq i}\wp '(x_i-x_j).
$$
The matrices $A, B,C$ are off-diagonal while the matrices $D, D'$ are diagonal.
The equation of the spectral curve is $\det \Bigl (L-(3z^2 +6\alpha_1)I\Bigr )=0$.

The linear system (\ref{a1}) is overdetermined. 
Differentiating the first equation in (\ref{a1}) with respect to $t$, we see that
the compatibility condition of the linear problems (\ref{a1}) is 
\beq\label{a2}
\Bigl (\dot L+[L,M]\Bigr ) {\bf c} =0.
\eeq
One can prove the following matrix identity (see the appendix):
\beq\label{a3}
\dot L+[L,M]=-12 D'\Bigl (L-(3z^2+6\alpha_1)I\Bigr )-\ddot X +12D'(6D-\dot X)+6\dot D -6D''',
\eeq
where $\displaystyle{D'''_{ik}=\delta_{ik}\sum_{j\neq i}\wp '''(x_i-x_j)}$.
It then follows that the compatibility condition (\ref{a2}) is equivalent 
to vanishing of all elements of the diagonal matrix
$$
-\ddot X +12D'(6D-\dot X)+6\dot D -6D'''.
$$
This gives equations of motion for the poles $x_i$. Writing the diagonal elements
explicitly, we get:
$$
\ddot x_i +6\sum_{j\neq i}(\dot x_i +\dot x_j)\wp '(x_i-x_j)-72
\sum_{j\neq i}\sum_{k\neq i} \wp (x_i-x_j)\wp '(x_i-x_k)+6\sum_{j\neq i}
\wp '''(x_i-x_j)=0.
$$
Taking into account the identity $\wp '''(x)=12\wp (x)\wp '(x)$, we obtain the
equations of motion (\ref{int8}):
\beq\label{int8a}
\ddot x_i +6\sum_{j\neq i}(\dot x_i +\dot x_j)\wp '(x_i-x_j)-72\!\!
\sum_{j\neq k \neq i}\wp (x_i-x_j)\wp '(x_i-x_k)=0.
\eeq
The rational limit  
(when $\wp (x) \to 1/x^2$) reads
\beq\label{a4}
\ddot x_i -12 \sum_{j\neq i}\frac{\dot x_i+\dot x_j}{(x_i-x_j)^3}+
144 \! \sum_{j\neq k\neq i}\frac{1}{(x_i-x_j)^2(x_i-x_k)^3}=0.
\eeq

\section{Integrals of motion}

The Lax representation of equations (\ref{int8}) is missing. Instead of it, we have
the matrix relation
\beq\label{a5}
\dot L+[L,M]=-12 D'\Bigl (L-(3z^2+6\alpha_1)I\Bigr )
\eeq
equivalent to the equations of motion. This is a sort of the Manakov's triple
representation \cite{Manakov}.
This relation means that in contrast to the KP case, where we have the Lax
equation for the Lax matrix of the elliptic Calogero-Moser system,
eigenvalues of our ``Lax matrix'' $L$ are not conserved and the evolution
$L\to L(t)$ is not isospectral. Nevertheless, 
the equation of the spectral curve, $\det \Bigl (L-(3z^2 +6\alpha_1)I\Bigr )=0$,
is an integral of motion. Indeed, 
$$
\frac{d}{dt}\, \log \det \Bigl (L-(3z^2 +6\alpha_1)I\Bigr )=
\frac{d}{dt}\, \mbox{tr}\log \Bigl (L-(3z^2 +6\alpha_1)I\Bigr )
$$
$$
=\, \mbox{tr}\Bigl [ \dot L\Bigl (L-(3z^2 +6\alpha_1)I\Bigr )^{-1}\Bigr ]=
-12\, \mbox{tr}D' =0,
$$
where we have used relation (\ref{a5}) and the fact that 
$\displaystyle{\mbox{tr}\, D'=\sum_{i\neq j}\wp '(x_i-x_j)=0}$ ($\wp '$ is an odd function).
We recall that $\alpha_1=-\frac{1}{2}\, \wp (\lambda )$.
The expression 
$$
R(z, \lambda )=\det \Bigl (3(z^2 -\wp (\lambda ))I-L\Bigr )
$$ 
is a polynomial in $z$
of degree $2N$. Its coefficients are integrals of motion (some of them may be trivial).

The matrix
$L=L(z, \lambda )$, which has essential singularities at $\lambda =0$, can be 
represented in the form $L=G\tilde L G^{-1}$, where $\tilde L$ does not have 
essential singularities and $G$ is the diagonal matrix $G_{ij}=\delta_{ij}
e^{-\zeta (\lambda )x_i}$. Therefore, 
$$
R(z, \lambda )=\sum_{k=0}^{2N}R_k(\lambda )z^k,
$$
where the coefficients $R_k(\lambda )$ are elliptic functions of $\lambda$ with poles
at $\lambda =0$.

Let us give some examples. At $N=2$ we have
$$
\det \Bigl (3(z^2 -\wp (\lambda ))I-L\Bigr )
=9z^4+3z^2\Bigl (\dot x_1 +\dot x_2 -18\wp (\lambda )\Bigr )-36 z\wp '(\lambda )
-3\wp (\lambda )(\dot x_1 +\dot x_2)
$$
$$\phantom{aaaaaaaaa}
+\dot x_1 \dot x_2 -6(\dot x_1 +\dot x_2)\wp (x_1-x_2)-27 \wp ^2(\lambda )+9g_2,
$$
where $g_2$ is the coefficient in the expansion of the $\wp$-function near $x=0$:
$\wp (x)=x^{-2}+\frac{1}{20}\, g_2 x^2 +\frac{1}{28}\, g_3 x^4 +O(x^6)$. 
Therefore, in this case we have 
two integrals of motion: $I_1=\dot x_1 +\dot x_2$, $I_2=
\frac{1}{2}(\dot x_1^2 + \dot x_2^2) +6(\dot x_1 +\dot x_2)\wp (x_1-x_2)$. 

At $N=3$ a non-trivial calculation leads to the following result:
$$
\det \Bigl (3(z^2 -\wp (\lambda ))I-L\Bigr )=27z^6+9\Bigl (I_1-45\wp (\lambda )\Bigr )z^4
-540\wp '(\lambda )z^3
$$
$$
+\left [ \frac{3}{2}\, I_1^2-3I_2-54\wp (\lambda )I_1
-1215\wp ^2(\lambda )+243g_2\right ]z^2
-36\wp '(\lambda )\Bigl (I_1+9\wp (\lambda )\Bigr )z
$$
$$
+I_3-I_1I_2 +\frac{1}{6}I_1^3 +3\wp (\lambda )\Bigl (I_2 -\frac{1}{2}I_1^2\Bigr )
-27\wp ^2 (\lambda )I_1
+9g_2 I_1 -135 \wp ^3 (\lambda )-27 g_2 \wp (\lambda ) +216 g_3,
$$
where
\beq\label{a6}
\begin{array}{lll}
I_1&=&\dot x_1 +\dot x_2 +\dot x_3,
\\ &&\\
I_2&=&\frac{1}{2}\Bigl (\dot x_1^2+\dot x_2^2+\dot x_3^2\Bigr )
+6\dot x_1 \Bigl (\wp (x_{12})+\wp (x_{13})\Bigr )+
6\dot x_2 \Bigl (\wp (x_{21})+\wp (x_{23})\Bigr )
\\ && \\
&&
+\, 6\dot x_3 \Bigl (\wp (x_{31})+\wp (x_{32})\Bigr )
-36 \Bigl (\wp (x_{12})\wp (x_{13})+\wp (x_{12})\wp (x_{23})
+\wp (x_{13})\wp (x_{23})\Bigr ),
\\&&\\
I_3&=&\frac{1}{3}\Bigl (\dot x_1^3+\dot x_2^3+\dot x_3^3\Bigr )
+6\dot x_1^2 \Bigl (\wp (x_{12})+\wp (x_{13})\Bigr )+
6\dot x_2^2 \Bigl (\wp (x_{21})+\wp (x_{23})\Bigr )
\\&&\\
&&+\, 6\dot x_3^2 \Bigl (\wp (x_{31})+\wp (x_{32})\Bigr )
+12 \dot x_1\dot x_2 \wp (x_{12})+12 \dot x_1\dot x_3 \wp (x_{13})+
12 \dot x_2\dot x_3 \wp (x_{23})
\\ &&\\
&& -\, 864 \wp (x_{12})\wp (x_{13}) \wp (x_{23})
\end{array}
\eeq
are integrals of motion
(here $x_{ik}\equiv x_i-x_k$).

In general, we can prove that the following quantities are integrals of motion:
\beq\label{a7}
\begin{array}{l}
\displaystyle{I_1=\sum_{i}\dot x_i},
\\ \\
\displaystyle{I_2=\frac{1}{2}\sum_i \dot x_i^2+6 \sum_{i\neq j}\dot x_i \wp (x_{ij})
-18 \! \! \sum_{i\neq j\neq k}\wp (x_{ij})\wp (x_{ik})}.
\end{array}
\eeq
In the expression for $I_2$ the last sum is taken over all triples of distinct
numbers $i,j,k$ from $1$ to $N$. The conservation of $I_1$ means that 
the center of masses moves
uniformly, i.e., $\displaystyle{\sum_{i}\ddot x_i=0}$. 

For the prove that $\dot I_1=0$ we write, using equations of motion (\ref{int8})
and permuting the summation indices,
$$
\dot I_1=\sum_{i}\ddot x_i=72\! \sum_{i\neq j\neq k}\wp (x_{ij})\wp ' (x_{ik})
$$
$$
=\, 12 \! \sum_{i\neq j\neq k}\Bigl (\wp (x_{ij})\wp '(x_{ik})+\wp ' (x_{ij})\wp (x_{ik})
$$
$$
\phantom{aaaaaaaaaaa}+\, \wp (x_{ji})\wp '(x_{jk})+\wp ' (x_{ji})\wp (x_{jk})
$$
$$
\phantom{aaaaaaaaaaaa}+\, \wp (x_{ki})\wp '(x_{kj})+\wp '(x_{ki})\wp (x_{kj})\Bigr )
$$
$$
=\, 12 \! \sum_{i\neq j\neq k}\Bigl [\p_{x_i}\Bigl (\wp (x_{ij})\wp (x_{ik})\Bigr )
+\p_{x_j}\Bigl (\wp (x_{ji})\wp (x_{jk})\Bigr )
+\p_{x_k}\Bigl (\wp (x_{ki})\wp (x_{kj})\Bigr )\Bigr ]=0,
$$
where we have used the identity
\beq\label{a8}
\p_{x_i}\! \Bigl (\wp (x_{ij})\wp (x_{ik})\Bigr )
+\p_{x_j}\! \Bigl (\wp (x_{ji})\wp (x_{jk})\Bigr )
+\p_{x_k}\! \Bigl (\wp (x_{ki})\wp (x_{kj})\Bigr )=0.
\eeq
It is in fact equivalent to the well known identity
$$
\left |\begin{array}{ccc}
1& \wp (x_{ij})&\wp '(x_{ij})
\\ 
1& \wp (x_{jk})&\wp '(x_{jk})
\\ 
1& \wp (x_{ki})&\wp '(x_{ki})
\end{array}
\right |=0
$$
and can be proved by expanding near the possible poles at $x_i=x_j$ and $x_i=x_k$.

For the proof that $\dot I_2=0$ we write:
$$
\dot I_2=\sum_i \dot x_i \ddot x_i +6\sum_{i\neq j} \ddot x_i \wp (x_{ij})
+6 \sum_{i\neq j} \dot x_i (\dot x_i-\dot x_j)\wp '(x_{ij})-36\!\!
\sum_{i\neq j\neq k}(\dot x_i-\dot x_j)\wp '(x_{ij})\wp (x_{ik}).
$$
Substituting the equations of motion, we have:
$$
\dot I_2=-6\sum_{i\neq j}\dot x_i(\dot x_i+\dot x_j)\wp '(x_{ij})+72 \!\!
\sum_{i\neq j\neq k}\dot x_i \wp (x_{ij})\wp '(x_{ik})
$$
$$
-36 \sum_{i\neq j}\sum_{k\neq i} (\dot x_i+\dot x_k)\wp (x_{ij})\wp '(x_{ik})
+432\sum_{i\neq l}\! \sum_{j\neq k\neq i}\wp (x_{ij})\wp '(x_{ik})\wp (x_{il})
$$
$$
+6\sum_{i\neq j}\dot x_i (\dot x_i-\dot x_j)\wp '(x_{ij})-36\!\!
\sum_{i\neq j\neq k}(\dot x_i-\dot x_j)\wp '(x_{ij})\wp (x_{ik}).
$$
The terms containing velocities cancel automatically (taking into account that
$\wp '(x_{ij})=-\wp '(x_{ji})$) and we are left with
$$
\dot I_2=432\sum_{i\neq l}\! \sum_{j\neq k\neq i}\wp (x_{ij})\wp '(x_{ik})\wp (x_{il})
$$
$$
=36 \!\! \sum_{i\neq j\neq k\neq l}\Bigl [
\wp '(x_{ij})\wp (x_{ik})\wp (x_{il})+\wp (x_{ij})\wp '(x_{ik})\wp (x_{il})+
\wp (x_{ij})\wp (x_{ik})\wp '(x_{il})
$$
$$ \phantom{aaaaaaaaaaa}
+\wp '(x_{ji})\wp (x_{jk})\wp (x_{jl})+\wp (x_{ji})\wp '(x_{jk})\wp (x_{jl})+
\wp (x_{ji})\wp (x_{jk})\wp ' (x_{jl})
$$
$$ \phantom{aaaaaaaaaaa}
+\wp '(x_{ki})\wp (x_{kj})\wp (x_{kl})+\wp (x_{ki})\wp ' (x_{kj})\wp (x_{kl})+
\wp (x_{ki})\wp (x_{kj})\wp '(x_{kl})
$$
$$ \phantom{aaaaaaaaaaa}
+\wp '(x_{li})\wp (x_{lj})\wp (x_{lk})+\wp (x_{li})\wp '(x_{lj})\wp (x_{lk})+
\wp (x_{li})\wp (x_{lj})\wp '(x_{lk})\Bigr ]
$$
$$
+72\!\! \sum_{i\neq j\neq k}
\Bigl [\wp (x_{ij})\wp '(x_{ik})\wp (x_{ik})+\wp '(x_{ik})\wp ^2(x_{ij})-
\wp '(x_{ik})\wp (x_{kj})\wp (x_{ik})-\wp '(x_{ik})\wp ^2(x_{kj})
$$
$$ \phantom{aaaaaaaaa}
+\wp (x_{ik})\wp '(x_{ij})\wp (x_{ij})+\wp '(x_{ij})\wp ^2(x_{ik})-
\wp '(x_{ij})\wp (x_{kj})\wp (x_{ij})-\wp '(x_{ij})\wp ^2(x_{kj})
$$
$$ \phantom{aaaaaaaaa}
+\wp (x_{ij})\wp '(x_{jk})\wp (x_{jk})+\wp '(x_{jk})\wp ^2(x_{ij})-
\wp '(x_{jk})\wp (x_{ki})\wp (x_{jk})-\wp '(x_{jk})\wp ^2(x_{ki})\Bigr ]
$$
$$
=36 \!\! \sum_{i\neq j\neq k\neq l}\Bigl [
\p_{x_i}\! \Bigl (\wp (x_{ij})\wp (x_{ik}) \wp (x_{il})\Bigr )+
\p_{x_j}\! \Bigl (\wp (x_{ji})\wp (x_{jk}) \wp (x_{jl})\Bigr )
$$
$$ \phantom{aaaaaaaaaaaaaaaaa}
+\, \p_{x_k}\! \Bigl (\wp (x_{ki})\wp (x_{kj}) \wp (x_{kl})\Bigr )
+\p_{x_l}\! \Bigl (\wp (x_{li})\wp (x_{lj}) \wp (x_{lk})\Bigr )\Bigr ]
$$
$$
+72\!\!\! \sum_{i\neq j\neq k}\!\! 
\Bigl (\wp (x_{ij})\! +\! \wp (x_{jk})\! +\! \wp (x_{ki})\Bigr )
\Bigl [\p_{x_i}\! \Bigl (\wp (x_{ij})\wp (x_{ik})\Bigr )
+\p_{x_j}\! \Bigl (\wp (x_{ji})\wp (x_{jk})\Bigr )
+\, \p_{x_k}\! \Bigl (\wp (x_{ki})\wp (x_{kj})\Bigr )\Bigr ],
$$
where we permuted the summation indices and separated the terms with 
$l=j$ and $l=k$.
The last line vanishes because of identity (\ref{a8}). The rest also vanishes
due to the identity
\beq\label{a9}
\begin{array}{l}
\p_{x_i}\! \Bigl (\wp (x_{ij})\wp (x_{ik}) \wp (x_{il})\Bigr )+
\p_{x_j}\! \Bigl (\wp (x_{ji})\wp (x_{jk}) \wp (x_{jl})\Bigr )
\\ \\
\phantom{aaaaaaaaaaa}+ \p_{x_k}\! \Bigl (\wp (x_{ki})\wp (x_{kj}) \wp (x_{kl})\Bigr )
+\p_{x_l}\! \Bigl (\wp (x_{li})\wp (x_{lj}) \wp (x_{lk})\Bigr )=0.
\end{array}
\eeq
The proof of this identity is standard. The left hand side 
is an elliptic function of $x_i$. Expanding it near the possible poles at
$x_i=x_j$, $x_i=x_k$, $x_i=x_l$ one can see that it is regular, so it is a constant
independent of $x_i$. By symmetry, this constant does not depend also 
on $x_j, x_k$ and $x_l$. To see that this constant is actually zero, one can put
$x_i=x$, $x_j=2x$, $x_k=3x$, $x_l=4x$.

Another integral of motion for any $N$ is 
\beq\label{a10}
J=\det_{1\leq i,j\leq N}\Bigl [\delta_{ij}\dot x_i -6\delta_{ij}
\sum_{k\neq i}\wp (x_{ik})-6(1-\delta_{ij})\wp (x_{ij})\Bigr ].
\eeq
The conservation of $J$ follows from the fact that $\displaystyle{J=
\lim_{\lambda \to 0}R(\lambda ^{-1}, \lambda )}$. Indeed,
since $$\Phi '(x, \lambda )=\Phi (x, \lambda )
\Bigl (\zeta (x+\lambda )-\zeta (x)-\zeta (\lambda )\Bigr )$$ and
$$
\tilde \Phi (x, \lambda )=e^{\zeta (\lambda )x}\Phi (x, \lambda )=
\lambda ^{-1}+\zeta (x)+\frac{1}{2}\, \frac{\sigma ''(x)}{\sigma (x)}\, \lambda +
O(\lambda ^2),
$$
we have
$$
\tilde L(z, \lambda )=(z-\lambda^{-1})Y(z, \lambda )+\dot X -6D -6Q +O(\lambda ),
$$
where $Q$ is the matrix with matrix elements $Q_{ij}=(1-\delta_{ij})\wp (x_{ij})$
and $Y(z, \lambda )$ is a matrix which is regular at $z=\lambda^{-1}$. Therefore,
$R(\lambda^{-1}, \lambda )=\det (\dot X -6D -6Q) +O(\lambda )$.

\section{The spectral curve}

The equation of the spectral curve is
\beq\label{s1}
R(z, \lambda )=\det \Bigl (3(z^2-\wp (\lambda ))I-L(z, \lambda )\Bigr )=0.
\eeq
It is easy to see that $L(-z, -\lambda )=L^{T}(z, \lambda )$, so the spectral curve 
admits the involution $\iota : (z, \lambda )\to (-z, -\lambda )$. 
We have $\displaystyle{R(z, \lambda )=\sum_{k=0}^{2N}R_k(\lambda )z^k}$, where
$R_k(\lambda )$ are elliptic functions of $\lambda$
such that $R_k(-\lambda )=(-1)^k R_k (\lambda )$. The functions 
$R_k (\lambda )$ can be represented as linear combinations of $\wp$-function and
its derivatives. Coefficients
of this expansion are integrals of motion (see examples for $N=2$ 
and $N=3$ in the previous section). Fixing values of these integrals, we obtain
via the equation $R(z, \lambda )=0$ the algebraic curve $\Gamma$ which is a 
$2N$-sheet covering of the initial elliptic curve ${\cal E}$ realized as a factor
of the complex plane with respect to the lattice generated by $2\omega$, $2\omega '$.

In a neighborhood of $\lambda =0$   
the matrix $\tilde L$
can be written as
$$
\tilde L=-6\lambda ^{-1}(z-\lambda^{-1})(E-I)-6(z-\lambda^{-1})S+O(1),
$$ 
where
$E$ is the rank $1$ matrix with matrix elements $E_{ij}=1$ for all $i,j =1, \ldots , N$
and $S$ is the antisymmetric matrix with matrix elements $S_{ij}=\zeta (x_i-x_j)$,
$i\neq j$, $S_{ii}=0$.

Therefore, near $\lambda =0$ the function $R(z, \lambda )$ can be represented in the form
$$
R(z, \lambda )=\det \Bigl (3(z^2-\lambda^{-2})I+6\lambda^{-1}(z-\lambda^{-1})(E-I)+6
(z-\lambda^{-1})S+O(1)\Bigr )
$$
$$
=\, \det \Bigl (3(z-\lambda^{-1})^2I +6\lambda^{-1}(z-\lambda^{-1})E
+6(z-\lambda^{-1})S+O(1)\Bigr )
$$
$$
=\, 3^N (z-\lambda^{-1})^{2N}\det \left (I+\frac{2}{z\lambda -1}\, E +
\frac{2\lambda}{z\lambda -1}\, S +O(\lambda^2)\right ).
$$
Using the fact that $\det \Bigl (A+\varepsilon B\Bigr )=\det A 
\Bigl (1+\varepsilon \, \mbox{tr}\, (A^{-1}B)\Bigr )+O(\varepsilon ^2)$ for any two
matrices $A$, $B$ and the relation 
$(I-\alpha E)^{-1}=I+\frac{\alpha}{1-N\alpha}\, E$, we find
$$
\det \left (I+\frac{2}{z\lambda -1}\, E +
\frac{2\lambda}{z\lambda -1}\, S +O(\lambda^2)\right )
$$
$$
=\, \det \left (I+\frac{2}{z\lambda -1}\, E+O(\lambda^2) \right )
\left (1+\frac{2\lambda}{z\lambda -1}\, \mbox{tr} \, \Bigl (S -
\frac{2}{z\lambda \! +\! 2N\! -\! 1}\, ES\Bigr )+O(\lambda^2) \right ).
$$
But for any antisymmetric matrix $S$ $\mbox{tr}\, S =\mbox{tr}\, (ES)=0$, so we are left with
$$
R(z, \lambda )=3^N (z-\lambda^{-1})^{2N}\det \left (I+\frac{2}{z\lambda -1}\, E +
O(\lambda ^2)\right ).
$$
The matrix $E$ has eigenvalue $0$ with multiplicity $N-1$ and another eigenvalue 
equal to $N$. Therefore, we can write $R(z, \lambda )$ in the form
\beq\label{s2}
R(z, \lambda )=3^N \Bigl (z+(2N-1)\lambda^{-1}-f_{2N}(\lambda )\Bigr )
\Bigl (z-\lambda^{-1}-f_{1}(\lambda )\Bigr )
\prod_{i=2}^{2N-1}\Bigl (z-\lambda^{-1}-f_{i}(\lambda )\Bigr ),
\eeq
where $f_i$ are regular functions of $\lambda$ at $\lambda =0$. 
The involution $\iota$ implies that $f_{2N}$ and $f_1$ 
are odd functions: $f_{2N}(-\lambda )=-f_{2N}(\lambda )$, 
$f_{1}(-\lambda )=-f_{1}(\lambda )$
and the other sheets can be 
numbered in such a way that 
$f_i (-\lambda )=-f_{2N+1-i}(\lambda )$,
$i=2, 3, \ldots , N$.
This means that the function $z$
has simple poles on all sheets at the points $P_j$ ($j=1, \ldots , 2N$) located 
above $\lambda =0$. Its expansion in the
local parameter $\lambda$ on the sheets near these points is given by the multipliers
in the right hand side of (\ref{s2}).  So we have the following expansions 
of the function $z$ near 
the ``points at infinity'' $P_j$:
\beq\label{s2a}
\begin{array}{l}
z=\, \lambda^{-1}+f_j(\lambda ) \quad \mbox{near $P_j$}, \quad j=1, \ldots , 2N-1,
\\ \\
z=-(2N\! -\! 1)\lambda ^{-1}+f_{2N}(\lambda ) \quad \mbox{near $P_{2N}$}.
\end{array}
\eeq
Similarly to the spectral curve of the elliptic
Calogero-Moser model \cite{Krichever80}, one of the sheets is distinguished, as it can be seen 
from (\ref{s2}). We call it the upper sheet. There is also another distinguished sheet,
where the point $P_1$ is located (and where the corresponding function $f_1$ is odd).
We call it the lower sheet for brevity. The points $P_1, P_{2N}$ are two 
fixed points of the involution $\iota$.

Let us find genus $g$ 
of the spectral curve $\Gamma$. Applying the Riemann-Hurwitz formula
to the covering $\Gamma \to {\cal E}$, we have $2g-2=\nu$, where $\nu$ is the number 
of ramification points of the covering. The ramification points are zeros on $\Gamma$
of the function $\p R/\p z$. Differentiating equation (\ref{s2}) with respect to $z$, 
we can see that the function $\p R/\p z$ has simple poles at the points $P_j$ 
($j=1, \ldots , 2N-1$) on all
sheets except the upper one, where it has a pole of order $2N-1$. The number
of poles of any meromorphic function is equal to the number of zeros. Therefore,
$\nu = 2(2N-1)$ and so $g=2N$.

The spectral curve $\Gamma$ is not smooth because in general position the genus
of the curve which is a $2N$-sheet covering of an elliptic curve is
$g=N(2N-1)+1$.

\section{Analytic properties of the $\psi$-function on the spectral curve}

Let $P$ be a point of the curve $\Gamma$, i.e. $P=(z, \lambda )$, where 
$z$ and $\lambda$ are connected by the equation $R(z, \lambda )=0$. The coefficients
$c_i$ in the pole ansatz for the function $\psi$, after normalization, 
are functions on the curve $\Gamma$:
$c_i=c_i(t, P)$. Let us normalize them by the condition $c_1(0, P)=1$.
In fact the non-normalized components $c_i(0, P)$ are equal to 
$\Delta_i (0, P)$, where $\Delta_i (0, P)$ are suitable minors of the matrix
$3(z^2 -\wp (\lambda ))I-L(0)$. They are holomorphic functions on $\Gamma$ outside the
points above $\lambda =0$. After normalizing the first component, all other components
$c_i(0,P)$ become meromorphic functions on $\Gamma$ outside the points $P_j$ located
above $\lambda =0$. Their poles are zeros on $\Gamma$ of the first minor of the matrix
$3(z^2 -\wp (\lambda ))I-L(0)$, i.e., they are given by common solutions of 
equation (\ref{s1}) and the equation 
$\det \Bigl (3(z^2 -\wp (\lambda ))\delta_{ij}-L_{ij}(0)\Bigr )=0$, 
$i,j=2, \ldots , N$. The location of these poles depends on the initial data.

On all sheets except the lower one the leading term of the matrix $\tilde L$ 
as $\lambda \to 0$ is proportional to $E-I$. 
Finding explicitly eigenvectors of the matrix $E-I$, one can see that
in a neighborhood of the ``points at infinity'' $P_j$ ($j=2, \ldots , 2N$) the functions 
$c_i (0, P)$ have the form
\beq\label{s3}
c_i (0, P)=\Bigl (c_i^{0(j)} +O(\lambda )\Bigr )e^{-\zeta (\lambda )(x_i(0)-x_1(0))},
\quad 2\leq i\leq N, \quad j=2, \ldots 2N-1,
\eeq
where $\displaystyle{\sum_{i=2}^{N}c_i^{0(j)}=-1}$ and
\beq\label{s4}
c_i (0, P)=\Bigl (1 +O(\lambda )\Bigr )e^{-\zeta (\lambda )(x_i(0)-x_1(0))},
\quad 2\leq i\leq N, \quad j= 2N
\eeq
(on the upper sheet). On the lower sheet, the leading term of the matrix $\tilde L$ 
as $\lambda \to 0$ is $O(1)$. Expanding the matrix $\tilde L$ in powers of $\lambda$, we
have
$$
\Lambda I-\tilde L=6f_1'(0)E+\dot X -6D -6Q +O(\lambda ),
$$
where $Q$ is the matrix with matrix elements $Q_{ij}=(1-\delta_{ij})\wp (x_i-x_j)$.
Let $c_i^{0(1)}$ be the eigenvector of the matrix in the right hand side 
(taken at $t=0$) with zero eigenvalue normalized by the condition $c_1^{0(1)}=1$,
then in a neighborhood of the point $P_1$ we can write
\beq\label{s3a}
c_i (0, P)=\Bigl (c_i^{0(1)} +O(\lambda )\Bigr )e^{-\zeta (\lambda )(x_i(0)-x_1(0))},
\quad 2\leq i\leq N, \quad j=1.
\eeq

The fundamental matrix ${\cal S}(t)$ of solutions to the equation 
$\p_t {\cal S} =M{\cal S}$, ${\cal S}(0)=I$, is a regular function of 
$z, \lambda$ for $\lambda \neq 0$.
Using equation (\ref{a5})
(which plays the role of the Lax equation for our system), we can write
$$
\Bigl (\dot L +[L,M]+12D' (L-\Lambda I)\Bigr ){\bf c}(t)=0,
$$
where $\Lambda = 3(z^2 -\wp (\lambda ))$.
Substituting ${\bf c}(t)={\cal S}(t){\bf c}(0)$ and $M=\dot {\cal S} {\cal S}^{-1}$, 
we can rewrite this equation
as
$$
\Bigl [\p_t \Bigl ({\cal S}^{-1}(L-\Lambda I){\cal S}\Bigr ) +
12{\cal S}^{-1}D' (L-\Lambda I){\cal S}\Bigr ]{\bf c}(0)=0
$$
or, equivalently, in the form of the differential equation
$$
\p_t {\bf b}(t)=W(t){\bf b}(t), \qquad W(t)=12{\cal S}^{-1}D'{\cal S},
$$
for the vector
$
{\bf b}(t)={\cal S}^{-1}(L-\Lambda I){\bf c}(t)$ with the initial condition ${\bf b}(0)=0$.
This differential equation with zero initial condition has the unique solution
${\bf b}(t)=0$ for all $t>0$.
Therefore, since the matrix ${\cal S}$ is non-degenerate, 
it then follows that ${\bf c}(t)={\cal S}(t){\bf c}(0)$ is the 
common solution of the equations $\dot {\bf c}=M{\bf c}$ and $L{\bf c}=\Lambda {\bf c}$.
Thus the vector ${\bf c}(t, P)$ has the same $t$-independent poles as the vector
${\bf c}(0,P)$.

In order to find $c_i (t, P)$ near the pre-images of the point $\lambda =0$ it is convenient
to pass to the gauge equivalent pair $\tilde L$, $\tilde M$, where
$$
\tilde L=G^{-1}LG , \quad \tilde M =-G^{-1}\p_t G +G^{-1}MG
$$
with the same diagonal matrix $G$ as before. Let $\tilde {\bf c}=G^{-1}{\bf c}$ be the
gauge-transformed vector ${\bf c}=(c_1, \ldots , c_N)^T$, then our linear system is
$$
\tilde L \tilde {\bf c}=3(z^2\! -\! \wp (\lambda ))\tilde {\bf c}, \qquad
\p_t \tilde {\bf c}=\tilde M \tilde {\bf c}.
$$

By a straightforward calculation one can check that the following relation holds:
\beq\label{s5}
\tilde M=-\lambda^{-1}\tilde L+(3z\lambda^{-2}\! -\! 4\lambda^{-3})I+
6(z-\lambda^{-1})(Q-D) +O(1).
\eeq
(It should be taken into account that $z$ is of order $O(\lambda^{-1})$, see (\ref{s2a}),
so the terms proportional to $z$ have to be kept in the expansion.)
Applying the both sides to an eigenvector $\tilde {\bf c}$ of $\tilde L$
with the eigenvalue $3(z^2\! -\! \wp (\lambda ))=3(z^2-\lambda^{-2})+O(\lambda^2)$, we get
\beq\label{s6}
\p_t \tilde {\bf c}=-z^3\tilde {\bf c}+(z-\lambda^{-1})^3\tilde {\bf c}
+6(z-\lambda^{-1})(Q-D)\tilde {\bf c} +O(1).
\eeq
Therefore, since $z=\lambda^{-1}+O(1)$ on all sheets except the upper one, we have
\beq\label{s7}
\p_t \tilde {\bf c}^{(j)}=-(z^3+O(1))\tilde {\bf c}^{(j)}, \quad j=1, \ldots , 2N-1,
\eeq
so 
$$
\tilde {\bf c}^{(j)}(t,P)=({\bf c}^{0(j)}+O(\lambda ))
e^{-z^3t}, \quad j=1, \ldots , 2N-1.
$$
In order to find the time dependence of the vector $\tilde {\bf c}^{(2N)}$ on the
upper sheet, we note that the corresponding eigenvector of the matrix $\tilde L$
is proportional to the vector ${\bf e}=(1, 1, \ldots , 1)^T$ 
with an addition of terms of order $O(1)$ and also note that $(Q-D){\bf e}=0$.
Therefore, since $z=-(2N-1)\lambda^{-1}+f_{2N}$ on the upper sheet, we have
\beq\label{s7a}
\p_t \tilde {\bf c}^{(2N)}=\Bigl 
(-z^3+k^3(\lambda ) +O(1)\Bigr )\tilde {\bf c}^{(2N)}, 
\eeq
where
$$
k(\lambda )=-2N\lambda^{-1}+f_{2N},
$$
so
$$
\tilde {\bf c}^{(2N)}(t,P)=({\bf e}+O(\lambda ))e^{(-z^3+k^3(\lambda ))t} .
$$
Coming back to the vector ${\bf c}(t,P)$, we obtain after normalization
\beq\label{s8}
c_i^{(j)}(t,P)=c_{ij}(\lambda ) e^{-\zeta (\lambda )(x_i(t)-x_1(0))+\nu_j (\lambda )t},
\eeq
where $\nu_j =-z^3$ for $j=1, \ldots , 2N-1$, $\nu_{2N}=-z^3 +k^3 (\lambda )$ and
$c_{ij}(\lambda )$ are regular functions in a neighborhood of $\lambda =0$. Their
values at $\lambda =0$ are
\beq\label{s9}
c_{1j}(0)=1, \quad j=1, \ldots , 2N, \qquad c_{ij}(0)=c_i^{0(j)}, \quad i\geq 2, \,\,
j\neq 2N, \qquad c_{i\, 2N}(0)=1,
\eeq
with $\displaystyle{\sum_{i=2}^{N}c_i^{0(j)}=-1}$ for $j=2, \ldots , 2N-1$.

After investigating the analytic properties of the vector ${\bf c}(t,P)$ 
let us turn to the function $\psi$:
$$
\psi (x, t, P)=\sum_{i=1}^{N}c_i(t, P)\Phi (x-x_i , \lambda )e^{zx+z^3t}.
$$
The function $\Phi (x-x_i, \lambda )$ has essential singularities at all points
$P_j$ located above $\lambda =0$. It follows from (\ref{s8}) that in the function
$\psi$ these essential 
singularities cancel on all sheets except the upper one, where 
$\psi \propto e^{k(\lambda )x+k^3(\lambda )t}e^{\zeta (\lambda )x_1(0)}$. 
From (\ref{s9}) it follows that
$\psi$ has simple poles at the points $P_1, P_{2N}$ (the two fixed points
of the involution $\iota$) and no poles at the points 
$P_j$ for $j=2, \ldots 2N-1$. The residue at the pole at $P_1$ is constant 
as a function of $x, t$. This is in agreement with the fact that the differential
operators $B_3, B_5$ (\ref{int3}) have no free terms, and so the result of 
their action to a 
constant vanishes.

The function $\psi$ also has other poles in the
finite part of the curve $\Gamma$, which do not depend on $x,t$.
Presumably, their number is $2N-2$ but the argument which allows one to 
count the number of poles of the $\psi$-function 
in the KP case (see \cite{Krichever80}) does not work for BKP.

\section{Conclusion}

In this paper we have derived equations of motion for poles of double-periodic
(elliptic) solutions to the BKP equation (equations (\ref{int8})). In contrast to the 
equations of motion for poles of elliptic solutions to the KP equation, where
interaction between ``particles'' (poles) is pairwise, in the BKP case
there are both two-body and three-body interactions. To the best of our knowledge,
many-body integrable systems with three-body interaction were never mentioned
in the literature (see, however, \cite{Wolfes74,CM74}, where some three-body integrable 
systems with three-body interaction were discussed). Instead of the Lax representation, 
the equations of motion admit a kind of the Manakov's triple representation. 

There are some problems which require further investigation. 
First, the Hamiltonian structure
of equations (\ref{int8}) is not known.
Besides, since the Lax representation is missing, integrability of equations (\ref{int8})
is not clear. Nevertheless, we believe that the system is integrable since the equation
of the spectral curve depending on the spectral parameter provides a large supply 
of independent conserved quantities. Three of them are known explicitly for any $N$. 
Another problem is to complete the proof that the $\psi$-function (\ref{ba1}) is the 
Baker-Akiezer function on the spectral curve. To do that, one should invent a way to 
count the number of poles of the $\psi$-function in the finite part of the spectral curve.

\section*{Appendix}
\def\theequation{A\arabic{equation}}
\setcounter{equation}{0}

\subsection*{Proof of equation (\ref{a3})}

Here we prove the main identity (\ref{a3}).
Using the explicit form of the matrices $L$, $M$ (\ref{a1a}), (\ref{a1b}), we write
$$
\begin{array}{lll}
\dot L+[L,M]&=&36z^2 \Bigl ([A,B]+[A,D]\Bigr )
\\ && \\
&&-\, 6z \Bigl (\dot A -[\dot X, B]\Bigr )
\\ &&\\
&&+\, 36 z\Bigl ([A,C]-[A, D']+2[B, D]\Bigr )
\\ &&\\
&&-\, 6 \Bigl (\dot B -[\dot X, C]\Bigr )-\ddot X +6\dot D
\\ &&\\
&&+\, 36 \Bigl ([B,C]-[B, D']+[C, D]\Bigr ).
\end{array}
$$
First of all we notice that $\dot A_{ik}=(\dot x_i-\dot x_k)\Phi '(x_i-x_k)$,
$\dot B_{ik}=(\dot x_i-\dot x_k)\Phi ''(x_i-x_k)$, and, therefore, 
we have $\dot A =[\dot X, B]$, $\dot B =[\dot X, C]$. To transform the commutators
$[A,B]+[A,D]$, we use the identity
\beq\label{A1}
\Phi (x )\Phi '(y)-\Phi (y)\Phi '(x)=\Phi (x+y)(\wp (x) -\wp (y))
\eeq
which, in turn, follows from the easily proved identity
\beq\label{A2}
\Phi (x, \lambda )\Phi (y, \lambda )=\Phi (x+y, \lambda )
\Bigl (\zeta (x)+\zeta (y)-\zeta (x+y+\lambda )+\zeta (\lambda )\Bigr ).
\eeq
With the help of (\ref{A1}) we get for $i\neq k$
$$
\Bigl ([A,B]+[A,D]\Bigr )_{ik}
$$
$$
=\, \sum_{j\neq i,k}\Phi (x_i-x_j)\Phi '(x_j-x_k)-
\sum_{j\neq i,k}\Phi ' (x_i-x_j)\Phi (x_j-x_k)
$$
$$
+\, \Phi (x_i-x_k)\Bigl (\sum_{j\neq k}\wp (x_j-x_k)-\sum_{j\neq i}\wp (x_i-x_j)\Bigr )=0,
$$
so $[A,B]+[A,D]$ is a diagonal matrix. To find its matrix elements, we use the limit
of (\ref{A1}) at $y=-x$:
\beq\label{A3}
\Phi (x)\Phi '(-x)-\Phi (-x)\Phi '(x)=\wp '(x)
\eeq
which leads to
$$
\Bigl ([A,B]+[A,D]\Bigr )_{ii}
$$
$$
=\,
\sum_{j\neq i}\Bigl (\Phi (x_i-x_j)\Phi ' (x_j-x_i)-\Phi ' (x_i-x_j)\Phi (x_j-x_i)\Bigr )
=\sum_{j\neq i}\wp '(x_i-x_j)=D'_{ii},
$$
so we finally obtain the matrix identity
\beq\label{A4}
[A,B]+[A,D]=D'.
\eeq

Combining the derivatives of (\ref{A1}) w.r.t. $x$ and $y$, we obtain the 
identities
\beq\label{A5}
\Phi (x)\Phi ''(y)-\Phi (y)\Phi ''(x)=2\Phi '(x+y)(\wp (x)-\wp (y))
+\Phi (x+y)(\wp '(x)-\wp '(y)),
\eeq
\beq\label{A6}
\Phi '(x)\Phi ''(y)-\Phi '(y)\Phi ''(x)=\Phi ''(x+y)(\wp (x)-\wp (y))
+\Phi '(x+y)(\wp '(x)-\wp '(y)).
\eeq
Their limits as $y\to -x$ are
\beq\label{A7}
\Phi (x)\Phi ''(-x)-\Phi (-x)\Phi ''(x)=0,
\eeq
\beq\label{A8}
\Phi '(x)\Phi ''(-x)-\Phi '(-x)\Phi ''(x)=-\frac{1}{6}\, \wp '''(x)+2\alpha_1\wp '(x).
\eeq
Using these formulas, it is easy to prove the following matrix identities:
\beq\label{A9}
[A,C]=2[D, B]+D'A+AD',
\eeq
\beq\label{A10}
[B,C]=[D,C]+D'B+BD'-\frac{1}{6}\, D''' +2\alpha_1 D'
\eeq
which are used to transform $\dot L+[L,M]$ to the form (\ref{a3}).

\subsection*{Some useful identities}

Apart from already mentioned identities for the $\Phi$-function for the calculations 
in Section 3 we need 
the following ones:
\beq\label{A11}
\Phi (x)\Phi (-x)=\wp (\lambda )-\wp (x),
\eeq
\beq\label{A12}
\Phi '(x)\Phi (-x)+\Phi '(-x)\Phi (x)=\wp '(\lambda ),
\eeq
\beq\label{A13}
\Phi '(x)\Phi ' (-x)=\wp ^2(x)+\wp (\lambda )\wp (x)+\wp ^2(\lambda )-
\frac{1}{4}\, g_2,
\eeq
\beq\label{A14}
\Phi (x)\Phi '' (-x)=\wp ^2(\lambda )+\wp (\lambda )\wp (x)-2\wp ^2(x ),
\eeq
\beq\label{A15}
\Phi '(x)\Phi '' (-x)=\Bigl (\wp '(\lambda )-\wp '(x)\Bigr )
\Bigl (\wp (x)+\frac{1}{2}\, \wp (\lambda )\Bigr ).
\eeq
They eventually follow from the basic identity (\ref{A2}). We also need some 
identities for the Weierstrass functions:
\beq\label{A16}
2\zeta (\lambda )-\zeta (\lambda +x)-\zeta (\lambda -x)=
\frac{\wp '(\lambda )}{\wp (x)-\wp (\lambda )},
\eeq
\beq\label{A16a}
\wp '^2(x)=4\wp ^3(x)-g_2\wp (x)-g_3,
\eeq
\beq\label{A17}
\wp (x+\lambda )-\wp (x-\lambda )=-\, \frac{\wp '(\lambda )\wp '(x)}{(\wp (x)-
\wp (\lambda ))^2},
\eeq
\beq\label{A18}
\wp (x+\lambda )+\wp (x-\lambda )=\frac{1}{2}\,
\frac{\wp '^2(x)+\wp '^2(\lambda )}{(\wp (x)-
\wp (\lambda ))^2}-2\Bigl (\wp (x)+\wp (\lambda )\Bigr ),
\eeq
\beq\label{A19}
\begin{array}{c}
2\wp (x)\Bigl (\wp (x-a)+\wp (a)+\wp (x)\Bigr )-\wp '(x)
\Bigl (\zeta (x-a)+\zeta (a)-\zeta (x)\Bigr )
\\ \\
=\, \wp (x)\wp (a)+\wp (x)\wp (x-a) +\wp (a)\wp (x-a) +\frac{1}{4}\, g_2.
\end{array}
\eeq
The last identity can be proved by expanding the both sides near the poles
at $x=0$ and $x=a$.

\section*{Acknowledgments}

One of the authors (A.Z.) thanks I. Krichever and S. Natanzon for useful discussions.
This research was carried out within the HSE University Basic 
Research Program and funded (jointly) by the Russian Academic Excellence Project '5-100'.
This work was also supported in part by RFBR grant
18-01-00461. The other author (D.R.) thanks Blokhin NMRCO, where part of the work was done, for their hospitality, and R. Klabbers for useful insights into Mathematica.


\begin{thebibliography}{99}



\bibitem{AMM77}
H. Airault, H.P. McKean, and J. Moser, {\it Rational and 
elliptic solutions of the
Korteweg-De Vries equation and a related many-body problem},
Commun. Pure Appl. Math., {\bf 30} (1977) 95--148.

\bibitem{Calogero71}
F. Calogero, {\it Solution of the one-dimensional
$N$-body problems with quadratic
and/or inversely quadratic pair potentials}, J. Math. Phys.
{\bf 12} (1971) 419–-436.

\bibitem{Calogero75} F. Calogero, {\it Exactly solvable one-dimensional many-body
systems}, Lett. Nuovo Cimento {\bf 13} (1975) 411--415.

\bibitem{Moser75}
J. Moser, {\it Three integrable Hamiltonian systems connected with isospectral
deformations}, Adv. Math. {\bf 16} (1975) 197--220.

\bibitem{Krichever78}
I.M. Krichever, {\it Rational solutions of the Kadomtsev-Petviashvili
equation and integrable systems of $N$ particles on a line},
Funct. Anal. Appl. {\bf 12:1} (1978) 59--61.

\bibitem{CC77} D.V. Chudnovsky, G.V. Chudnovsky, {\it Pole expansions of non-linear
partial differential equations}, Nuovo Cimento {\bf 40B} (1977) 339--350.

\bibitem{Krichever80} I.M. Krichever, {\it Elliptic solutions of the Kadomtsev-Petviashvili
equation and integrable systems of particles}, Funk. Anal. i Ego Pril. {\bf 14:4} (1980) 45--54
(in Russian); English translation:
Functional Analysis and Its Applications {\bf 14:4} (1980) 282–-290.

\bibitem{OP81} M.A. Olshanetsky and A.M. Perelomov, {\it Classical integrable
finite-dimensional systems related to Lie algebras}, Phys. Rep. {\bf 71} (1981) 313--400.

\bibitem{GH84} J. Gibbons and T. Hermsen,
{\it A generalization of the Calogero-Moser system},
Physica D {\bf 11} (1984) 337–-348.

\bibitem{RS86} S.N.M. Ruijsenaars and H. Schneider, {\it 
A new class of integrable systems and its relation to
solitons},
 Annals of Physics {\bf 146} (1986) 1--34.
 
 \bibitem{Ruij87} S.N.M. Ruijsenaars, {\it Complete integrability of relativistic
 Calogero-Moser systems and elliptic function identities}, 
 Commun. Math. Phys. {\bf 110} (1987) 191--213.
 
 \bibitem{KKS78} D. Kazhdan, B. Kostant and S. Sternberg, {\it Hamiltonian group
 actions and dynamical systems of Calogero type}, Comm. Pure Appl. Math. {\bf 31} (1978)
 481--507.
 
\bibitem{GN95} A. Gorsky and N. Nekrasov, {\it Relativistic Calogero-Moser
model as gauged WZW theory}, Nucl. Phys. {\bf B436} (1995) 582--608.

\bibitem{Nekr99} N. Nekrasov, {\it Infinite-dimensional algebras, many-body systems and
gauge theories}, In: Moscow Seminar in Mathematical Physics, AMS Transl. Ser. 2,
vol. 191 (1999) 263--299.

\bibitem{vanD94} J.F. van Diejen, {\it Integrability of difference Calogero-Moser 
systems}, J. Math. Phys. {\bf 35} (1994) 2983--3004.

\bibitem{Ch19} O. Chalykh, {\it Quantum Lax pairs via Dunkl and Cherednik 
operators}, Commun. Math. Phys. {\bf 369} (2019) 261--316.


\bibitem{KBBT95} I. Krichever, O. Babelon, E. Billey and M. Talon, {\it Spin generalization of the
Calogero-Moser system and the matrix KP equation}, Amer. Math. Soc. Transl. Ser. 2
{\bf 170} (1995) 83--119.


 
 \bibitem{KZ95} I. Krichever and A. Zabrodin, {\it 
Spin generalization of the Ruijsenaars-Schneider model, non-abelian 2D
Toda chain and representations of Sklyanin algebra}, Uspekhi Mat. Nauk
{\bf 50} (1995) 3--56 (in Russian) (English translation: 
Russ. Math. Surv., {\bf 50} (1995) 1101--1150).



\bibitem{DJKM83} E. Date, M. Jimbo, M. Kashiwara and T. Miwa,
{\it Transformation groups for soliton equations: Nonlinear integrable systems --
classical theory and quantum theory} (Kyoto, 1981). Singapore: World Scientific,
1983, 39--119.

\bibitem{DJKM82} E. Date, M. Jimbo, M. Kashiwara and T. Miwa,
{\it Transformation groups for soliton equations IV. A new hierarchy
of soliton equations of KP type}, Physica D {\bf 4D} (1982) 343--365.

\bibitem{DJKM82a} E. Date, M. Jimbo, M. Kashiwara and T. Miwa,
{\it Quasi-periodic solutions of the orthogonal KP equation.
Transformation groups for soliton equations V}, Publ. RIMS, Kyoto Univ.
{\bf 18} (1982) 1111--1119.

\bibitem{LW99} I. Loris and R. Willox, {\it Symmetry reductions of the BKP
hierarchy}, Journal of Mathematical Physics {\bf 40} (1999) 1420--1431.

\bibitem{Tu07} M.-H. Tu, {\it On the BKP Hierarchy: Additional Symmetries,
Fay Identity and Adler–-Shiota–-van Moerbeke Formula}, Letters in Mathematical Physics
{\bf 81} (2007) 93--105.

\bibitem{Manakov} S. Manakov, {\it Method of inverse scattering problem
and two-dimensional evolution equations}, Uspekhi Mat. Nauk {\bf 31} (1976)
245--246.

\bibitem{Wolfes74}
J. Wolfes, {\it On the three-body linear problem with three-body interaction}, 
J. Math. Phys. {\bf 15} (1974) 1420—-1424.

\bibitem{CM74}
F. Calogero and C. Marchioro, {\it Exact solution of a one-dimensional 
three-body scattering problem with two-body and/or three-body
inverse-square potentials}, J. Math. Phys. {\bf 15} (1974) 1425—-1430. 

\end{thebibliography}
\end{document}